\documentclass[lettersize,journal]{IEEEtran}
\usepackage{amsmath,amsfonts}
\usepackage{algorithmic}
\usepackage{algorithm}
\usepackage{array}
\usepackage[caption=false,font=normalsize,labelfont=sf,textfont=sf]{subfig}
\usepackage{textcomp}
\usepackage{stfloats}
\usepackage{xurl}
\usepackage{verbatim}
\usepackage{makecell}
\usepackage{graphicx}
\usepackage{multirow}
\usepackage{hyperref}
\usepackage{siunitx}
\usepackage{xfrac}
\usepackage{mathtools}
\usepackage{threeparttable} 
\usepackage{eurosym}

\usepackage[table,dvipsnames]{xcolor}
\usepackage{nicematrix}
\usepackage{booktabs} 

\usepackage[backend=biber,style=ieee,sorting=none]{biblatex}
\definecolor{inner}{RGB}{244,150,72}
\definecolor{indirect}{RGB}{110,136,230}
\definecolor{direct}{RGB}{187,240,129}

\usepackage{tikz}
\usetikzlibrary{decorations.pathreplacing,calc}

\begin{document}

\title{Multi-Day Stochastic Dispatch for Power Systems With Long-Duration Storage Using Diffusion-Generated Weather Forecasts}

\author{Leonard Göke
\thanks{Leonard Göke is with the
Reliability and Risk Engineering Laboratory, Institute of Energy and Process
Engineering, ETH Zurich, 8092 Zurich, Switzerland.}
}



\maketitle

\begin{abstract}
As power systems shift toward high shares of wind and solar, long-duration storage becomes pivotal for balancing multi-day deficits and surpluses of uncertain supply and demand. Current dispatch strategies for long-duration storage either assign a fixed value to stored energy or extend the modeled horizon using a single forecast. Stochastic system dispatch with a multi-day horizon for long-duration storage remains unexplored, and existing stochastic dispatch only considers day-ahead uncertainty or Markovian uncertainty beyond one week of perfect foresight. Therefore, we benchmark stochastic multi-day dispatch against deterministic alternatives in a rolling-horizon setup. To emulate probabilistic multi-day forecasts, we combine diffusion with a semi-Markov model of weather regimes and a logistic growth model for meteorological forecast errors. At high renewable shares, stochastic operation reduces thermal backup generation and consistently eliminates loss of load. The benefits decrease with smaller shares of renewables, less long-duration storage, or higher forecast skill. Overall, our results suggest that systems with high shares of variable renewables can substantially benefit from probabilistic multi-day forecasts and stochastic dispatch.
\end{abstract}


\section{Introduction}

Wind and solar energy are key to keeping global warming below \qty{2}{\celsius}, to securing energy supply for countries that depend on fossil fuel imports, and to meeting the energy demands of artificial intelligence. Even many projections with substantial firm generation from nuclear, geothermal, or fossil plants with carbon capture still predict wind and solar shares well above 60\% \cite{Clarke2022}. However, integrating wind and solar into power systems remains challenging. Their output fluctuates with the weather, and these fluctuations are uncertain because forecasts are imperfect and lose accuracy as the forecast horizon lengthens.

Balancing variable renewables depends on energy storage, and the required storage duration increases with the share of wind and solar. At a supply share of around 50\%, a few hours of storage suffice. Accordingly, many power systems near this share are seeing rapid growth in lithium-ion battery capacity \cite{Schill2020}. Between 50\% and 90\%, a range many countries are expected to reach in the next decade, multi-day fluctuations call for long-duration storage, such as flow batteries, compressed air, or pumped hydro storage \cite{Victoria2019}. Above 90\% renewable shares, seasonal and multi-year balancing requires storing renewable fuels, primarily to buffer fuel production against final demand rather than for reconversion to electricity \cite{Goeke2025}.

This reliance on storage fundamentally changes how the power system must operate. Storage dispatch must anticipate surpluses and deficits to decide when and where storage units charge or discharge. Since these surpluses and deficits are weather-dependent, system operation becomes a decision problem under uncertainty.

Instead of multi-day horizons, existing work on power system dispatch considers two different timescales of uncertainty \cite{Roald2023}:
\begin{enumerate}
\item \textbf{Optimal power flow and unit commitment} models focus on the grid operation and uncertainty in renewable supply only up to one or two days ahead. Over such short horizons, forecast errors are small, but grid dispatch requires immediate balancing of even minor deviations. To this end, these models use stochastic optimization to schedule thermal power plants, ensuring sufficient balancing power and reserves. More recent work also includes battery storage as a balancing option \cite{Li2016,Papavasiliou2018}.

\item In systems with a large share of hydropower, \textbf{hydrothermal operation} models manage uncertainty in weather-dependent water inflows up to several years ahead \cite{Helseth2023}. Because large reservoirs can store inflows, operation aims to minimize reliance on thermal backup plants by avoiding both reservoir depletion and spillage. Since reservoirs buffer short-term uncertainty, time steps typically range from a week to a month. Within each step, models assume perfect foresight; across steps, they represent the uncertainty of inflows as a Markov process, an assumption justified by the long time steps \cite{Pereira1991,Fullner2025}.
\end{enumerate}

The few previous studies with a multi-day timescale do not examine macro-scale power systems but rather single-node microgrids. \textcite{Aaslid2022} use 6-hour stages for the first 60 hours and then daily stages for a month to model a Norwegian microgrid with 80-hour hydrogen storage. They find that the benefits of stochastic dispatch increase with the reliance on variable renewables and storage. \textcite{Craparo2017} use a two-stage approach and forecasts of up to 72 hours to study a hypothetical military grid with a generic storage technology, but find no benefit from look-aheads beyond 33 hours.

Since the existing literature does not address multi-day stochastic dispatch at the macro scale, there is also a gap in corresponding forecasting methods. Probabilistic forecasts for power system dispatch cover demand and renewable supply but typically only one, sometimes two days ahead \cite{Konstantinou2023,Zhang2025,TerrenSerrano2026}. Medium-range weather forecasts exist, but not in a form usable for system-scale dispatch \cite{Hong2020}. Instead, their scope focuses on individual variables at single sites, such as generation from a specific wind farm, matching the scope of microgrid studies \cite{Taylor2009}. 

Against this background, this paper benchmarks multi-day stochastic dispatch against two deterministic alternatives: a short horizon with a fixed value for stored energy, and an extended horizon with a single forecast \cite{Dalvi2026}. To enable this benchmark, we introduce a diffusion-based model that emulates probabilistic multi-day forecasts and allows us to vary forecast skill.

The remainder of this paper is organized as follows: Section~\ref{sec:meth1} presents an approach to emulate probabilistic multi-day forecasts that combines a diffusion model with a semi-Markov model for weather regimes and a logistic growth model for meteorological forecast errors. Section~\ref{sec:meth2} describes the rolling-horizon setup used to benchmark stochastic dispatch against deterministic alternatives, and Section~\ref{sec:cs} introduces the numerical case study. Section~\ref{sec:res1} evaluates the accuracy of the forecast emulation, and Section~\ref{sec:res2} quantifies the benefit of the stochastic dispatch resulting from the rolling-horizon benchmark. Section~\ref{sec:cl} concludes the paper.

\section{Forecast emulation for residual load} \label{sec:meth1} 

This section introduces the forecast emulator, which enables the rolling-horizon analysis under multi-day uncertainty. The emulator directly forecasts residual load, the difference between electricity demand and renewable supply, which ultimately determines storage dispatch.

Rather than producing true real-time forecasts, the emulator generates plausible forecasts retrospectively, once the realized weather is known. This approach allows for treating forecast accuracy as an exogenous parameter, grounded in the meteorological limits on predictability. The alternative, developing a true multi-day forecasting method, would not only exceed the scope of a first study on this topic but also risk poor forecast quality and, thus, overestimate the benefits of stochastic dispatch.

\subsection{Auto-regressive diffusion for hourly profiles} \label{sec:diffMod} 

The core of the forecast emulator is a diffusion model that generates the raw hourly residual-load forecast $R^{\text{hr,raw}}_d$ for day $d$. The forecast covers all hours $H$ of the day and all zones $Z$ in normalized space: 
\begin{equation}  
R^{\text{hr,raw}}_d \in [-1,1]^{|Z| \times |H|} \label{eq:defR} 
\end{equation}

\begin{figure}[!t]
\centering
\includegraphics[scale=1.0]{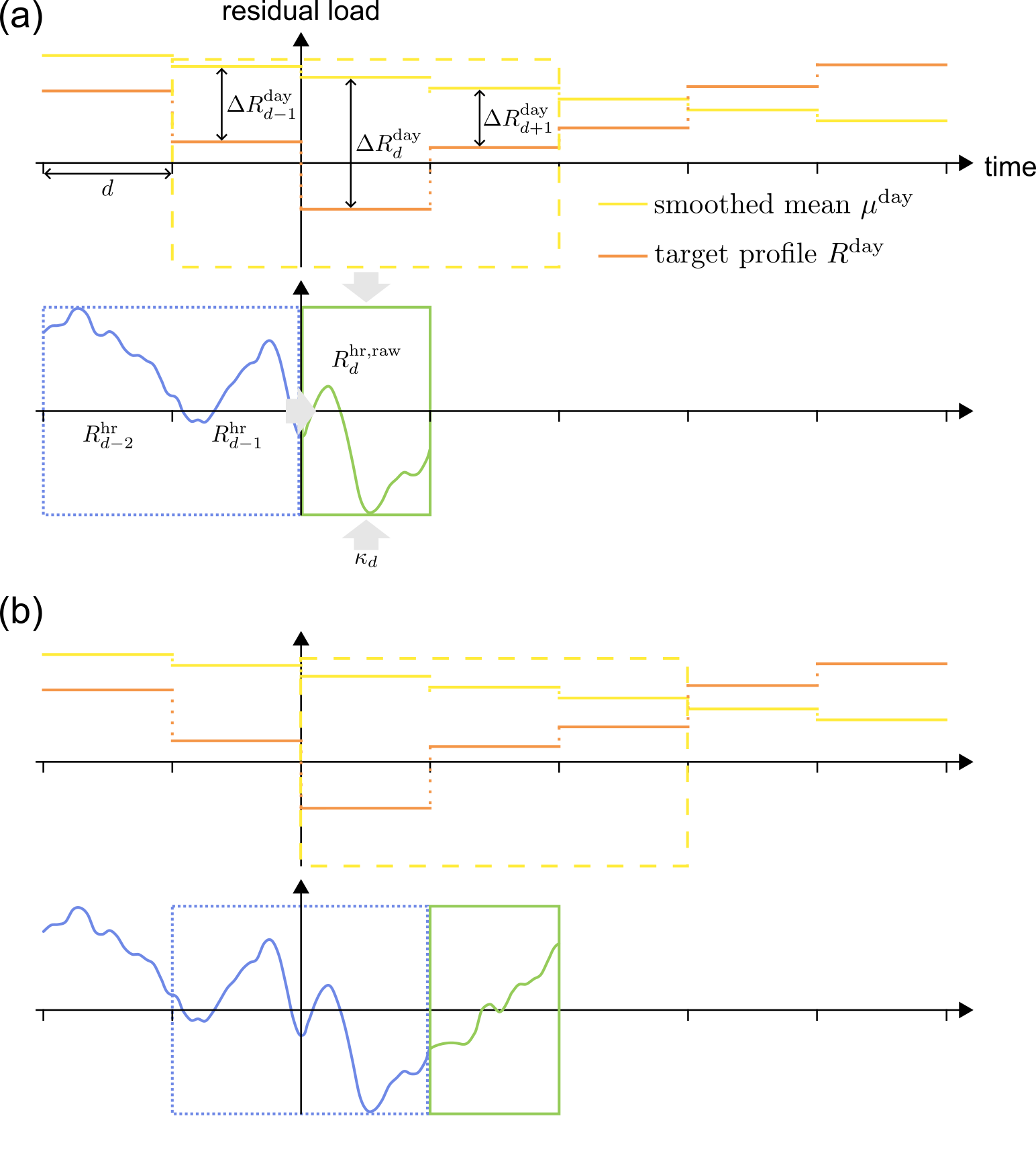}
\caption{Process and context information of the auto-regressive diffusion model.}
\label{fig:diff}
\end{figure}

Fig.~\ref{fig:diff} illustrates the diffusion process for a single zone. To enable a stochastic rolling-horizon setup, the model is auto-regressive, and its context includes the hourly residual load of the two preceding days, $R^{\text{hr}}_{d-2}$ and $R^{\text{hr}}_{d-1}$, which are the model's own previous outputs. Extending the context beyond two days did not noticeably reduce the validation loss during training.

The calendar context $\kappa_d$ in Eq.~\eqref{eq:kappa} provides two kinds of information. The first two elements encode the day of the year $\tilde{d}$ with a sine and a cosine. Using two trigonometric functions allows us to achieve continuity at the end of the year and clearly distinguish seasons. With one trigonometric function, spring and autumn, for instance, would have the same values. The remaining elements form a binary vector $w_d \in \{0,1\}^7$ that indicates the weekday, which matters for the demand component of residual load.

\begin{equation}
\kappa_d = \big[\sin(2\pi \tilde{d}/365),\, \cos(2\pi \tilde{d}/365),\, w_d^\top\big]^\top \in \mathbb{R}^{9} \label{eq:kappa}
\end{equation}

The final context element $\psi_d$ adds information on the daily residual load of the previous day $d-1$, the generated day $d$, and the next day $d+1$ in each zone. Because this context requires information on the realized time series in advance, this element makes the entire model a retrospective emulator rather than a true prediction. 

Instead of absolute levels, $\psi_d$ in Eq.~\eqref{eq:diff_level} uses the daily anomalies of residual load $\Delta R^{\text{day}}_{d}$:
\begin{equation}
\psi_d = \big[\,\Delta R^{\text{day}}_{d} - \Delta R^{\text{day}}_{d-1},\; \Delta R^{\text{day}}_{d},\; \Delta R^{\text{day}}_{d+1} - \Delta R^{\text{day}}_{d} \big]^\top \in \mathbb{R}^{3|Z|} \label{eq:diff_level}
\end{equation}

The anomaly in Eq.~\eqref{eq:ana} is the difference between the daily residual load $R^{\text{day}}_{d}$ and its smoothed mean $\mu^{\text{day}}_{d}$ over all training years. The smoothed mean is a harmonic regression with one sine/cosine pair and avoids sampling noise when working with a limited number of training years.
\begin{equation}
\Delta R^{\text{day}}_{d} = R^{\text{day}}_{d} - \mu^{\text{day}}_{d} \label{eq:ana}
\end{equation}
Using anomalies rather than absolute levels facilitates forecasting by separating the seasonal cycle from weather-driven deviations. The changes of the anomaly to the previous and next day in $\psi_d$ ensure a smooth hourly time series without implausible gradients around the turn of the day. Although the network only receives anomalies and never explicit absolute levels, the previous hourly profiles $R^{\text{hr}}_{d-2}$ and $R^{\text{hr}}_{d-1}$ and the calendar context $\kappa_d$ implicitly indicate the absolute residual load of the generated day.

With this context, the diffusion model can be expressed as the conditional distribution $p_\theta$ over all possible profiles $R^{\text{hr,raw}}_d$ in Eq.~\eqref{eq:diff_cond}. The distribution is not given explicitly but is defined by the denoising process: starting from Gaussian noise, a neural network $\varepsilon_\theta$ predicts and removes noise at each denoising step, with $\theta$ denoting the network weights learned during training.
\begin{equation}
R^{\text{hr,raw}}_d \sim p_\theta\big(R^{\text{hr,raw}}_d \,\big|\, \underbrace{R^{\text{hr}}_{d-2},\, R^{\text{hr}}_{d-1}}_{\text{history}},\,\underbrace{\kappa_d}_{\text{calendar}},\, \underbrace{\psi_d}_{\text{daily level}}\big) \label{eq:diff_cond}
\end{equation}

The auto-regressive setup of the diffusion model risks drifting towards implausible values that exceed the normalization interval. Conditioning on daily levels already mitigates this problem, but the raw profiles $R^{\text{hr,raw}}_d$ are additionally clipped. Afterward, a level shift in $\operatorname{artanh}$-space, described in Eq.~\eqref{eq:diff_shift}, moves the profiles towards the targeted daily residual load without leaving the normalization interval again. Within the $\tanh$, the first term represents the raw hourly values, and the difference between the second and third terms is the offset between the targeted and the current daily level. Due to the nonlinearity of $\tanh$, the scaling remains approximate.
\begin{equation}
\begin{split}
R^{\text{hr}}_{d,h,z} = {} & \tanh\!\Big( \operatorname{artanh}\big(R^{\text{hr,raw}}_{d,h,z}\big) \\
& {} + \operatorname{artanh}\big(\mu^{\text{day}}_{d,z} + \Delta R^{\text{day}}_{d,z}\big) \\
& {} - \operatorname{artanh}\Big(\tfrac{1}{|H|}\sum\limits_{h' \in H} R^{\text{hr,raw}}_{d,h',z}\Big) \Big) \; \forall \, h, z
\end{split} \label{eq:diff_shift}
\end{equation}
Implicitly, the clipping treats the extremes of residual load as inputs defined by the normalization interval rather than as outcomes of the forecast emulator.

\subsection{Semi-Markov model for daily weather regimes}  \label{sec:smm} 

As the next step, a semi-Markov model generates a time series of daily residual load anomalies, $\Delta \hat{R}^{\text{day}}_{d}$, in each zone. With this series as context, the diffusion model generates the realized hourly residual load time series, which the forecasts then use as their reference. 

The semi-Markov model describes fluctuations in residual load around its seasonal mean, capturing the duration and magnitude of energy surpluses and deficits. The process consists of different regimes, each with a fixed anomaly, but how long each regime lasts and which regime follows are drawn from probability distributions. Accordingly, the daily process remembers the duration of a regime and is not Markovian. Only the sequence of regime switches is, since the next regime depends only on the current one. The regime approach reflects the meteorological weather regimes that underlie fluctuations in residual load \cite{Bloomfield2020}.

In this study, the parameters of the semi-Markov model are month-specific because seasonal variations in electricity demand and renewable supply lead to strong heteroskedasticity in daily residual load anomalies. For each month $m$, the set $\mathcal{X}_m$ in Eq.~\eqref{eq:hsmm_input} contains the daily anomaly vectors across zones for all training days in that month, where $m(d)$ denotes the month of day $d$:
\begin{equation}
\mathcal{X}_m = \Big\{ \Delta R^{\text{day}}_d \in \mathbb{R}^{|Z|} \;\Big|\; d \in D^{\text{train}},\; m(d) = m \Big\} \label{eq:hsmm_input}
\end{equation}
The $k$-means algorithm assigns each of these days to a regime $g$ and determines the anomaly $\Delta \dot{R}^{\text{day}}_{m,g}$ of each regime as the centroid of its assigned days. The number of regimes per month is set at the elbow point of the within-cluster sum of squares.

Empirical probability distributions for durations and transitions result from assigning regimes to the training days of each month, as illustrated in Tables~\ref{tab:hsmm_example} and~\ref{tab:hsmm_params}. Table~\ref{tab:hsmm_example} shows exemplary anomaly data and the regime assigned by k-means clustering. Table~\ref{tab:hsmm_params} shows the empirical distribution of durations $P^{\text{dur}}$, i.e., the number of consecutive days in the same regime, and the empirical distribution of transitions $P^{\text{trn}}$ into the next regime $g'$.

\begin{table}[!t]
\centering
\caption{Illustrative Example of Regime Assignment}
\label{tab:hsmm_example}
\footnotesize
\renewcommand{\arraystretch}{1.1}
\setlength{\tabcolsep}{4pt}
\begin{tabular}{@{}lrrrc@{}}
\toprule
 & \multicolumn{3}{c}{$\Delta R^{\text{day}}$ (GW)} & \\
\cmidrule(lr){2-4}
Day & $z{=}1$ & $z{=}2$ & $z{=}3$ & Regime $g$ \\
\midrule
Jan 10 & $-2$  & $-4$  & $-2$ & 2 \\
Jan 11 & $-1$  & $-3$  & $-2$ & 2 \\
Jan 12 & $7$   & $6$   & $5$  & 3 \\
Jan 13 & $10$  & $8$   & $7$  & 3 \\
Jan 14 & $13$  & $12$  & $7$  & 3 \\
Jan 15 & $-4$  & $-4$  & $2$  & 2 \\
Jan 16 & $2$   & $-1$  & $-1$ & 2 \\
Jan 17 & $-9$  & $-10$ & $-6$ & 1 \\
Jan 18 & $-10$ & $-11$ & $-9$ & 1 \\
Jan 19 & $-9$  & $-10$ & $-7$ & 1 \\
Jan 20 & $-3$  & $-5$  & $-2$ & 2 \\
Jan 21 & $0$   & $-3$  & $-1$ & 2 \\
Jan 22 & $-4$  & $-5$  & $-2$ & 2 \\
Jan 23 & $8$   & $7$   & $6$  & 3 \\
Jan 24 & $7$   & $7$   & $5$  & 3 \\
Jan 25 & $-8$  & $-9$  & $-6$ & 1 \\
Jan 26 & $1$   & $-2$  & $-1$ & 2 \\
Jan 27 & $2$   & $0$   & $0$  & 2 \\
\bottomrule
\end{tabular}
\end{table}

\begin{table}[!t]
\centering
\caption{Semi-Markov Parameters Estimated From the Example in Table~\ref{tab:hsmm_example}}
\label{tab:hsmm_params}
\footnotesize
\renewcommand{\arraystretch}{1.1}
\setlength{\tabcolsep}{4pt}
\begin{tabular}{@{}crrrcccccc@{}}
\toprule
 & \multicolumn{3}{c}{$\Delta \dot{R}^{\text{day}}$} & \multicolumn{3}{c}{$P^{\text{dur}}$} & \multicolumn{3}{c}{$P^{\text{trn}}$} \\
\cmidrule(lr){2-4} \cmidrule(lr){5-7} \cmidrule(l){8-10}
$g$ & $z{=}1$ & $z{=}2$ & $z{=}3$ & $\ell{=}1$ & $\ell{=}2$ & $\ell{=}3$ & $g'{=}1$ & $g'{=}2$ & $g'{=}3$ \\
\midrule
1 & $-9$ & $-10$ & $-7$ & $\sfrac{1}{2}$ & 0 & $\sfrac{1}{2}$ & 0 & 1 & 0 \\
2 & $-1$ & $-3$  & $-1$ & 0 & $\sfrac{2}{3}$ & $\sfrac{1}{3}$ & $\sfrac{1}{3}$ & 0 & $\sfrac{2}{3}$ \\
3 & $9$  & $8$   & $6$  & 0 & $\sfrac{1}{2}$ & $\sfrac{1}{2}$ & $\sfrac{1}{2}$ & $\sfrac{1}{2}$ & 0 \\
\bottomrule
\end{tabular}
\end{table}

To continue regimes across month boundaries, the regimes of each month are ranked by their anomaly $\Delta \dot{R}^{\text{day}}$ summed across zones. At a month boundary, the ongoing regime continues as the regime of the same rank in the new month, thereby keeping the jump in the anomaly small. The duration of a run follows the distribution of the month in which it starts. In contrast, the probability of transitioning to the next regime depends on the new month.

On this basis, the semi-Markov model generates the time series of daily anomalies of the realized data over the days of the rolling-horizon analysis $D^{\text{roll}}$:

\begin{equation}
\big(\Delta \hat{R}^{\text{day}}_d\big)_{d \in D^{\text{roll}}} \sim \text{SMM}\big(\Delta \dot{R}^{\text{day}}_{m,g},\,  P_{m,g}^{\text{dur}},\, P_{m,g,g'}^{\text{trn}} \big) \label{eq:gen_smm}
\end{equation}

\subsection{Logistic error growth for forecast accuracy}

To generate plausible forecasts, the emulator adds an error to the time series of realized daily anomalies, which then serves as an alternative input to the diffusion model. Since this model is auto-regressive, forecasts start from the same history as the realized data; since it is stochastic, hourly forecasts deviate from the realized profiles even where the daily levels agree.

The error model builds on the fact that, due to the chaotic nature of atmospheric processes, the quality of a forecast deteriorates with lead time $\tau$ and converges logistically to climatological error. The climatological error is the error of a forecast that ignores initial conditions, i.e., the present state of the atmosphere, and builds solely on long-term distributions. Accordingly, empirical samples of climatological error in daily residual load are obtained by comparing the value for a specific calendar day and year with the average across all other training years for that day.

To describe the logistic error growth towards the climatology error, Eqs.~\eqref{eq:err_a} to~\eqref{eq:err_c} introduce the relative error $e$, the differential equation describing its growth, and the closed-form solution based on \textcite{Dalcher1987}. In the closed-form solution, $e_0$ refers to the initial error and $\alpha$ controls the speed of convergence.
\begin{alignat}{2}
e & = \frac{E}{E^{\text{clim}}} && \label{eq:err_a} \\
\frac{\mathrm{d}e}{\mathrm{d}\tau} & = \alpha \, e \, (1 - e) && \label{eq:err_b} \\
e(\tau) & = \frac{e_0}{e_0 + (1 - e_0) \cdot \exp(-\alpha \tau)} && \label{eq:err_c}
\end{alignat}

Two known points determine the two parameters of the relative error function. The first is the day-ahead error relative to the climatology error, $p^{\text{day}}$, which gives $e(1) = p^{\text{day}}$, since day-ahead errors are frequently reported for power systems. The second is an assumption on the lead time $\tau^{\text{clim}}$ after which a forecast has lost almost all of its value, reflected by a high relative error $p^{\text{clim}}$. On this basis, we can solve for $\alpha$ and $e_0$:
\begin{alignat}{2}
\alpha & = \frac{1}{\tau^{\text{clim}}-1}\left[\ln\frac{p^{\text{clim}}}{1-p^{\text{clim}}} - \ln\frac{p^{\text{day}}}{1-p^{\text{day}}}\right] \label{eq:err_alpha} \\
e_0 & = \frac{p^{\text{day}}}{p^{\text{day}} + (1-p^{\text{day}})\,\exp(\alpha)} \label{eq:err_e0}
\end{alignat}

\begin{figure}[!t]
\centering
\includegraphics[scale=1.0]{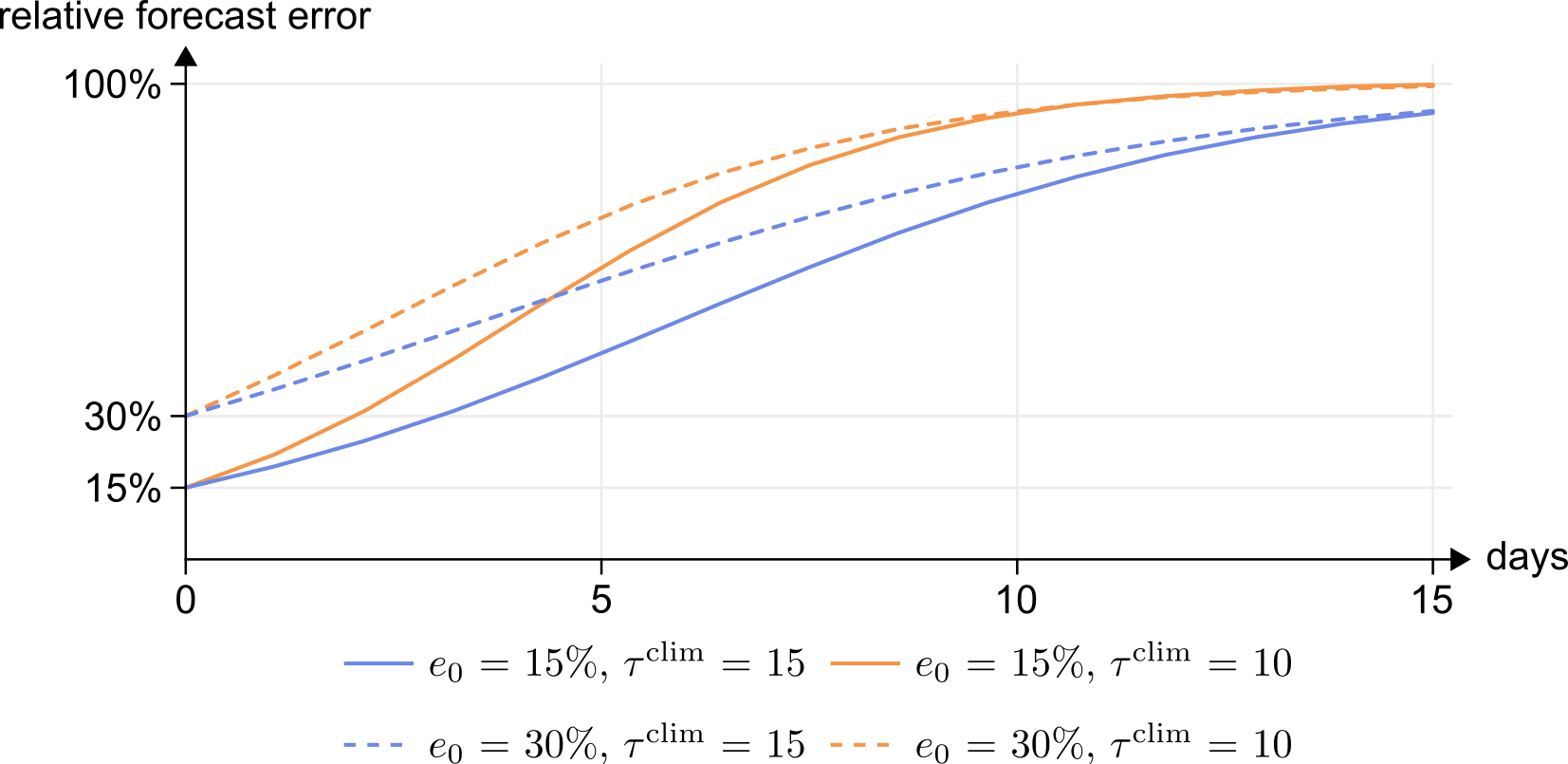}
\caption{Logistic growth of forecast errors depending on the initial error and the saturation time for a 95\% error.}
\label{fig:error}
\end{figure}

Fig.~\ref{fig:error} plots exemplary curves for the relative error. The case study section elaborates on the exact parametrization.

With the relative error function, we can compute the daily anomaly $\Delta \tilde{R}^{\text{day}}_{d,\tau}$ forecasted at the start of day $d$ with a lead time of $\tau$ days:
\begin{equation}
\begin{aligned}
\Delta \tilde{R}^{\text{day}}_{d,\tau} = {} & \Delta \hat{R}^{\text{day}}_{d+\tau-1} + s_{d,\tau} \cdot \boldsymbol{e}(\tau) \odot \big|E^{\text{clim}}_d\big| \\
& \forall \, \tau \in \{1, \dots, \delta \cdot \Lambda\}
\end{aligned} \label{eq:err_forecast}
\end{equation}
$E^{\text{clim}}_d$ is the climatology error drawn once per forecast from the empirical distribution described above. Since all zones are drawn from the same historical year, the spatial correlation of errors is preserved. The relative error $\boldsymbol{e}(\tau)$ is a vector of zone-specific error functions, because the day-ahead error $p^{\text{day}}$ depends on the composition of residual load in each zone; $\odot$ denotes the element-wise product.

The sign of the error $s_{d,\tau}$ is uniform across zones and chosen such that the forecast follows the weather dynamics of the historical year from which the climatology error was drawn. For instance, when creating a forecast on the 10th of December until the 24th of December, and the drawn climatology error is from the historical year 2010, $s_{d,\tau}$ is picked to minimize the difference in day-to-day changes of residual load across all zones between the time series from the 10th to the 24th of December 2010 and the generated forecast. This approach achieves a natural evolution of errors, preventing two equally unrealistic behaviors: a persistent over- or underestimation, or excessive jumps between over- and underestimation. The minimization is implemented as a small mixed-integer linear program.

Finally, the considered lead times $\tau$ correspond to the look-ahead of the rolling-horizon dispatch, which is given by the length per stage $\delta$ and the number of stages $\Lambda$. Again, the diffusion model computes hourly residual load profiles from the daily anomalies.

\section{Rolling-horizon dispatch} \label{sec:meth2}

This section first introduces the dispatch problem underlying the rolling-horizon setup and then the setup itself.

\subsection{Dispatch problem}

The optimization problem in Eq.~\eqref{eq:tree}, used in the rolling-horizon dispatch and the perfect-foresight reference, optimizes operation across the nodes $N$, which are linked by the storage levels $l_n$ in each zone $z$ and for each storage technology $k$. Each node covers one stage of $\delta$ days, and the nodes are organized in a tree, with $a(n)$ denoting the unique ancestor of node $n$ and $N^{\text{end}}$ the nodes on the final stage. 

\begin{subequations}
\begin{alignat}{1}
\min_{l} \; & {\overbrace{\sum_{n \in N} \rho_n \cdot \varphi_n\big(l_{a(n)}, l_n\big)\vphantom{\big(W^\top\big)}}^{\text{operating costs}}} \, \, - {\overbrace{\sum_{n \in N^{\text{end}}} \rho_n \cdot W_{d+\delta \cdot \Lambda}^\top \, l_n}^{\text{value of final storage levels}}}  \label{eq:tree_a} \\
\text{s.t.} \; & l_{a(1)} = L_d  \label{eq:tree_c} \\
& l_n \in \mathbb{R}^{|Z| \cdot |K|}_{\geq 0}  \nonumber
\end{alignat} \label{eq:tree}
\end{subequations}

The objective includes the expected operating costs, computed from the probability $\rho_n$ and the operating costs $\varphi_n$ of each node, minus the expected end-of-horizon value of stored energy. According to the tree structure, the probabilities $\rho_n$ of all nodes on the same stage sum to one. The end-of-horizon value follows from the vector $W_{d+\delta \cdot \Lambda}$, which provides the zone- and technology-specific values of stored energy.

Eq.~\eqref{eq:disp} describes how the dispatch within each node determines its costs depending on the storage levels at its beginning, which correspond to the levels at the end of its ancestor $l_{a(n)}$, and at its own end $l_{n}$. The objective function minimizes the costs of thermal generation, the product of the specific costs $C^{\text{th}}$ and the generation $f$, plus the costs of loss of load, the product of the specific costs $C^{\text{lol}}$ and the loss of load $q$.

The energy balance in Eq.~\eqref{eq:disp_b} introduces the hourly residual load $R^{\text{hr}}_{n,t,z}$ from the previous section, converted from normalized to physical units. In the constraint, the residual load is covered by the sum of thermal generation $f$, loss of load $q$, storage discharging and charging, $s^{\text{out}}$ and $s^{\text{in}}$, and the net exchange $u$ with other zones. The dispatch problem uses a transport model for exchange, given the focus on large market zones, and omits unit commitment or ramping constraints, given the two-day perfect-foresight horizon used in the case study.

The other constraints ensure technical consistency: Eq.~\eqref{eq:disp_c} tracks the storage level $s^{\text{lvl}}$ within the node, accounting for self-discharge $\zeta$ and for charging and discharging losses $\eta$. Eq.~\eqref{eq:disp_d} fixes the storage levels to the prescribed initial and final levels. Eqs.~\eqref{eq:disp_e} to~\eqref{eq:disp_g} ensure that storage, thermal generation, and exchange comply with the installed capacities $X$.

\begin{figure*}[!t]
\normalsize
\begin{subequations}
\begin{alignat}{2}
\varphi_n\big(l_{a(n)}, l_{n}\big) := \min \; \; & \sum_{t \in T} \sum_{z \in Z} \big( C^{\text{th}} \cdot f_{t,z} + C^{\text{lol}} \cdot q_{t,z} \big) && \label{eq:disp_a} \\
\text{s.t.} \; \; & f_{t,z} + q_{t,z} + \sum_{k \in K} \big( s^{\text{out}}_{t,z,k} - s^{\text{in}}_{t,z,k} \big) && \nonumber \\
& \; \; + \sum_{z' \in Z \setminus \{z\}} \big( u_{t,z',z} - u_{t,z,z'} \big)\geq R^{\text{hr}}_{n,t,z} && \; \; \forall \, t \in T, z \in Z \label{eq:disp_b} \\
& s^{\text{lvl}}_{t,z,k} = \zeta_k \cdot s^{\text{lvl}}_{t-1,z,k} + \eta^{\text{in}}_k \cdot s^{\text{in}}_{t,z,k} - \frac{s^{\text{out}}_{t,z,k}}{\eta^{\text{out}}_k} && \; \; \forall \, t \in T, z \in Z, k \in K \label{eq:disp_c} \\
& s^{\text{lvl}}_{0,z,k} = l_{a(n),z,k}, \; s^{\text{lvl}}_{|T|,z,k} = l_{n,z,k} && \; \; \forall \, z \in Z, k \in K \label{eq:disp_d} \\
& s^{\text{in}}_{t,z,k} \leq X^{\text{in}}_{z,k}, \; s^{\text{out}}_{t,z,k} \leq X^{\text{out}}_{z,k}, \; s^{\text{lvl}}_{t,z,k} \leq X^{\text{size}}_{z,k} && \; \; \forall \, t \in T, z \in Z, k \in K \label{eq:disp_e} \\
& f_{t,z} \leq X^{\text{th}}_{z} && \; \; \forall \, t \in T, z \in Z \label{eq:disp_f} \\
& u_{t,z',z} \leq X^{\text{exc}}_{z',z} && \; \; \forall \, t \in T, (z',z) \in Z \times Z  \label{eq:disp_g} \\
& f_{t,z}, \, q_{t,z}, \, u_{t,z',z}, \, s^{\text{in}}_{t,z,k}, \, s^{\text{out}}_{t,z,k}, \, s^{\text{lvl}}_{t,z,k} \in \mathbb{R}_{\geq 0}&& \nonumber
\end{alignat} \label{eq:disp}
\end{subequations}
\hrulefill
\vspace*{4pt}
\end{figure*}

\subsection{Rolling-horizon setup}

\begin{figure}[!t]
\centering
\includegraphics[scale=1.0]{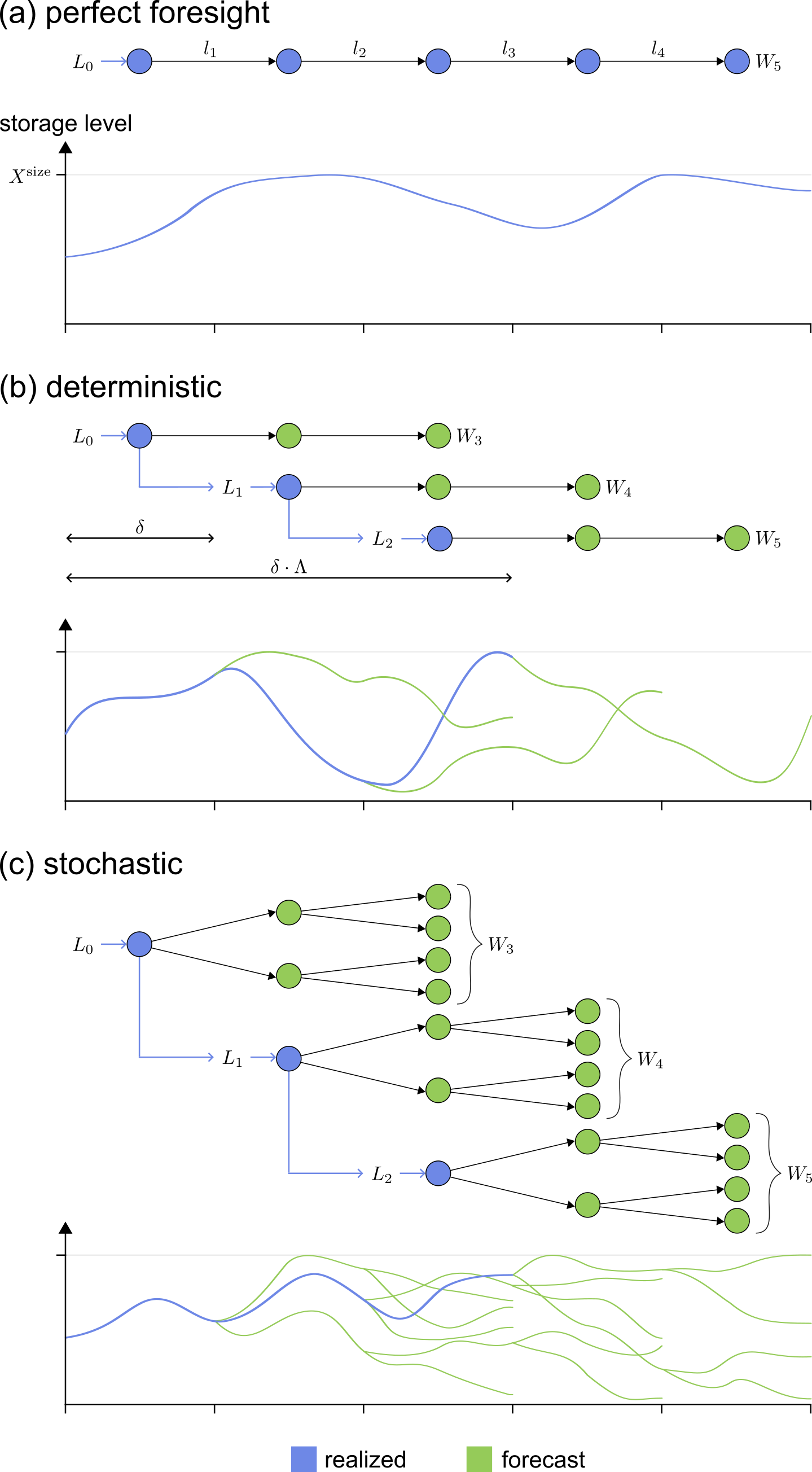}
\caption{Setups for the perfect-foresight reference and for deterministic and stochastic rolling-horizon dispatch.}
\label{fig:rolling}
\end{figure}

Building on this problem formulation, Fig.~\ref{fig:rolling} illustrates the setups for the perfect-foresight reference and for deterministic and stochastic rolling-horizon dispatch, using a simple example with five stages. In all cases, each stage has length $\delta$, and all setups ultimately run on the same realized data but differ in the anticipated future conditions.

The perfect-foresight setup describes a hypothetical case without uncertainty and provides a lower bound for the costs of the rolling-horizon setups. It is a single optimization based solely on the realized data, which perfectly anticipates future conditions and computes the optimal storage operation given the initial levels $L_0$ and the final values of stored energy $W_{5\delta}$.

The deterministic setup limits perfect foresight to the first stage and relies on a single forecast for the remaining $\Lambda - 1$ stages of the look-ahead horizon. Only the storage decisions for the first stage are final and determine the starting conditions for the next step of the rolling horizon, which reveals the realized data for the following stage, and so on. Stored energy at the end of the horizon is again assigned a corresponding value.

To generate the forecast for each step, we draw exactly one climatology error from the empirical distribution for the respective calendar day and compute the corresponding hourly time series of residual load as described in Eq.~\eqref{eq:err_forecast}.

The stochastic setup follows the same rolling-horizon approach as the deterministic setup, but uses a scenario tree rather than a single forecast to anticipate future conditions and capture the time persistence of weather \cite{Bloomfield2020}. Therefore, the first-stage decisions are no longer optimal for a single forecast, but must hedge against a broad range of scenarios weighted by their probability.

To generate the scenario tree of forecasts, we first compute an hourly time series of all errors in the empirical distribution, providing a large number of independent forecast paths. In the next step, we use the greedy algorithm described in \textcite{Pflug2014} to fit a scenario tree to this data. The algorithm starts with a single chain and repeatedly splits the path to reduce the error measured by k-medoids clustering until the node budget is exhausted.

\section{Case study} \label{sec:cs}

The case study runs the perfect-foresight reference and the deterministic and stochastic rolling-horizon dispatch on an 80-year time series with a stage length $\delta$ of two days. Preliminary tests showed that this horizon is necessary for the analyzed metrics to be robust, since loss of load is sensitive to rare but extreme deficit events. The perfect foresight over two days implied by the stage length complements existing work on uncertainty in power system dispatch, which either covers horizons of at most two days or assumes perfect foresight over at least one week. The number of nodes in the scenario tree for stochastic dispatch is set to 60.

The following subsections first describe the power system context for the case study and then the parametrization of the forecast emulator. All code and data to reproduce the results are openly available.\footnote{\url{https://github.com/leonardgoeke/multiDayStochasticDispatch/releases/tag/stochasticDispatchPaper}}

\subsection{Power system context}

The power system context for the study results from a separate capacity expansion problem that yields the installed capacities $X$, the end-of-horizon values for stored energy $W$, and the initial storage levels $L_0$. The capacity expansion problem uses the climate-robust formulation introduced in \cite{Goeke2025}. The zone- and technology-specific values for stored energy correspond to the dual variables of the storage balance, and the initial storage levels $L_0$ are the average storage levels at the beginning of the year.

To keep the computational complexity manageable, the case study includes three zones that loosely represent Switzerland, Germany, and France. For a consistent parametrization, capacity expansion and forecasting both use a European dataset with 35 years of climate data on electricity demand, capacity factors for wind, and capacity factors for solar \cite{pecd}. The transmission capacities $X^{\text{exc}}$ between zones are projections for 2040 by European grid operators \cite{entsoe2020}.

The capacity expansion model considers three generation technologies: wind, solar, and combined-cycle gas turbines (CCGT) as thermal backup. To achieve the targeted renewable share, CCGT generation is limited relative to final demand, for example, to 10\% in the 90\% renewable case. Within this limit, the capacity expansion model still determines CCGT capacity and, in turn, utilization. Wind and solar are subject to zone-specific expansion limits \cite{Auer2020}.

Finally, the capacity expansion considers two storage technologies: lithium-ion batteries and pumped-hydro storage, as a generic representation of LDS. The energy-to-power ratio is flexible, ranging from 1 to 4 hours for batteries and from 1 to 40 hours for pumped-hydro storage. Technical parameters and costs for all technologies are based on data from the Danish Energy Agency, which are widely used in European energy system studies \cite{DEA}.

\subsection{Forecast parametrization}

The empirical residual-load data for training the forecast emulator are derived from the wind and solar capacities computed by the capacity expansion model and the 35 years of climate data. Accordingly, the forecast emulator is updated for each capacity configuration.

As a first step, the hourly residual load is used to train the diffusion model described in Section~\ref{sec:diffMod}. Table~\ref{tab:diffusion} summarizes the settings of the diffusion model. The convolutional layers capture temporal patterns in the historical context, and separate encoders map the heterogeneous context inputs into a common embedding for the main network.

\begin{table}[!t]
\centering
\caption{Settings of the Diffusion Model}
\label{tab:diffusion}
\footnotesize
\renewcommand{\arraystretch}{1.1}
\begin{tabular}{@{}ll@{}}
\toprule
Component & Setting \\
\midrule
Noise schedule & 1000 steps, linear variance from $10^{-4}$ to $0.02$ \\
Sampling & 25 strided steps, stochastic \\
History encoder & 2 convolutional layers, 32 and 64 filters \\
Context encoders & Dense layers, 256 units \\
Main network & 3 dense layers, 512 units \\
\bottomrule
\end{tabular}
\end{table}

Next, the daily empirical residual load is used to estimate the semi-Markov model described in Section~\ref{sec:smm}. Based on an elbow analysis, the model uses 10 regimes per month. To generate the 80-year time series of daily residual load, we sample 200 candidate time series and select the one whose average residual load in each zone best matches the original data. This selection reduces random differences in residual load between runs with the same setup, thereby making the results comparable.

The last step, the error model, computes the climatological error from the daily empirical residual load. To compute the day-ahead error for the logistic error-growth function, we extract day-ahead forecasts and realized values of demand, wind generation, and solar generation from the database of European transmission system operators \cite{entsoeTransparency}. We then compute zone-specific relative errors for demand, wind, and solar and, assuming uncorrelated errors, aggregate them into an overall relative day-ahead error $p^{\text{day}}$ as a Euclidean sum weighted by energy quantities. For the climatological error, we follow the literature and assume that errors reach 95\% of their climatological level, $p^{\text{clim}}$, after a lead time $\tau^{\text{clim}}$ of 15 days \cite{Zhang2019, Dalcher1987}. Section~\ref{sec:res2b} performs a sensitivity analysis on this assumption.

\section{Accuracy of forecast emulation} \label{sec:res1}

This section verifies that the forecast emulator reliably reproduces the characteristics of the training data and provides plausible forecasts. All analyses in this section refer to the 90\% renewable case, whose wind and solar capacities are used to compute residual load.

\begin{figure}[!t]
\centering
\includegraphics[scale=1.0]{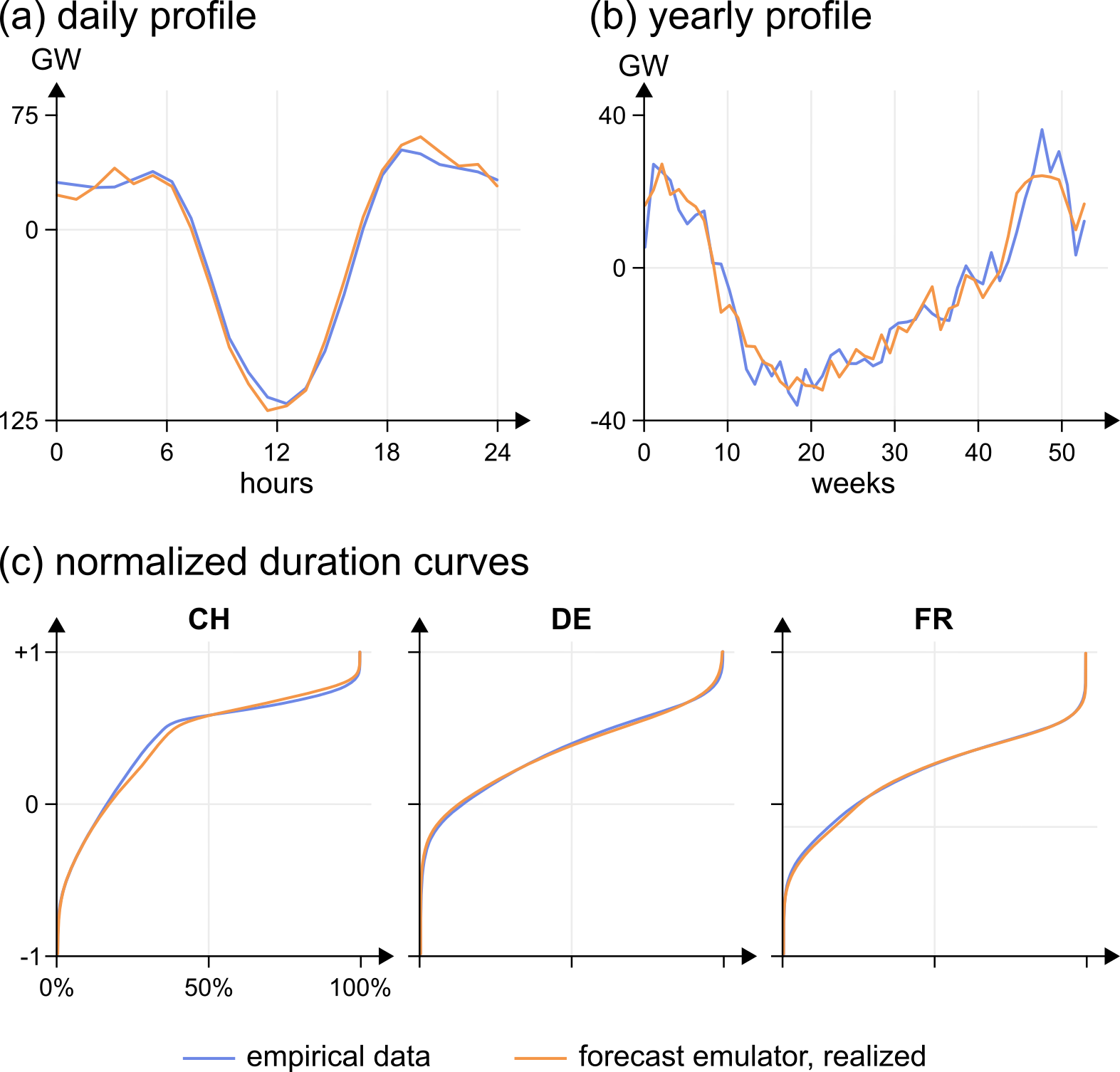}
\caption{Comparison of profiles between the training data and the output of the forecast emulator.}
\label{fig:acc2}
\end{figure}

The plots in Fig.~\ref{fig:acc2} demonstrate that the combination of the diffusion and semi-Markov models reproduces the daily profile, the yearly profile, and the residual-load duration curves in each zone. Small deviations in the daily profile likely result from rescaling the diffusion model's hourly output to align with the normalization interval and the targeted daily value. The yearly profiles still show noise, since the training data cover only 35 years and the generated data 80 years.

\begin{figure*}[!t]
\centering
\includegraphics[width=\textwidth]{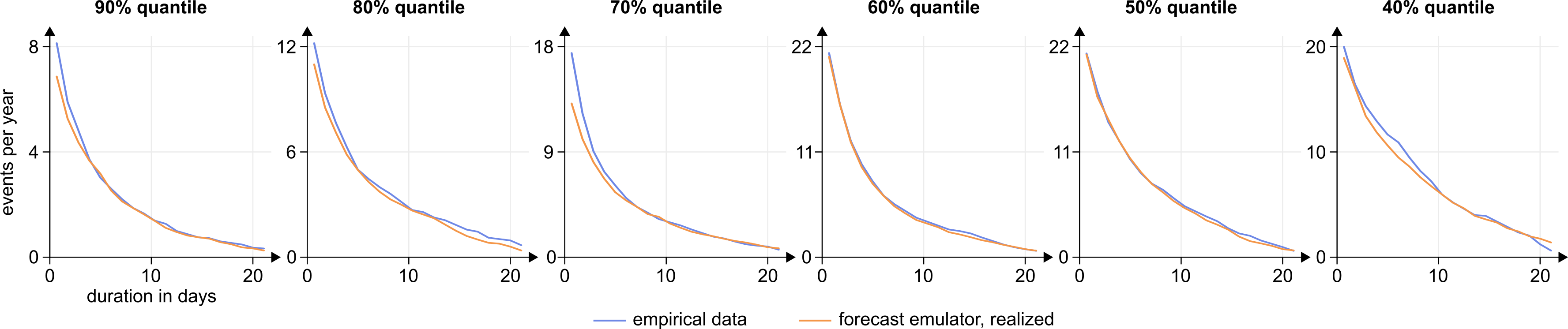}
\caption{Frequency, duration, and magnitude of deficit events according to the energy-drought metric of \cite{Kittel2026}.}
\label{fig:acc1}
\end{figure*}

Since our analysis focuses on LDS, Fig.~\ref{fig:acc1} checks whether the forecast emulator reproduces the frequency, duration, and magnitude of energy deficits in the underlying data. The figure adapts the energy-drought metric introduced in \cite{Kittel2026} by counting the number of periods in which the mean daily residual load exceeds a given threshold. For instance, the first curve shows how many periods per year with a mean residual load above its 90\% quantile last at least a given number of days. To avoid double-counting, the metric searches from longest to shortest duration, and days assigned to longer periods are excluded from shorter ones. Overall, the forecast emulator accurately reproduces the frequency and duration of these periods across thresholds, indicating that the combination of semi-Markov and diffusion model captures the persistence of residual load.

\begin{figure}[!t]
\centering
\includegraphics[scale=1.0]{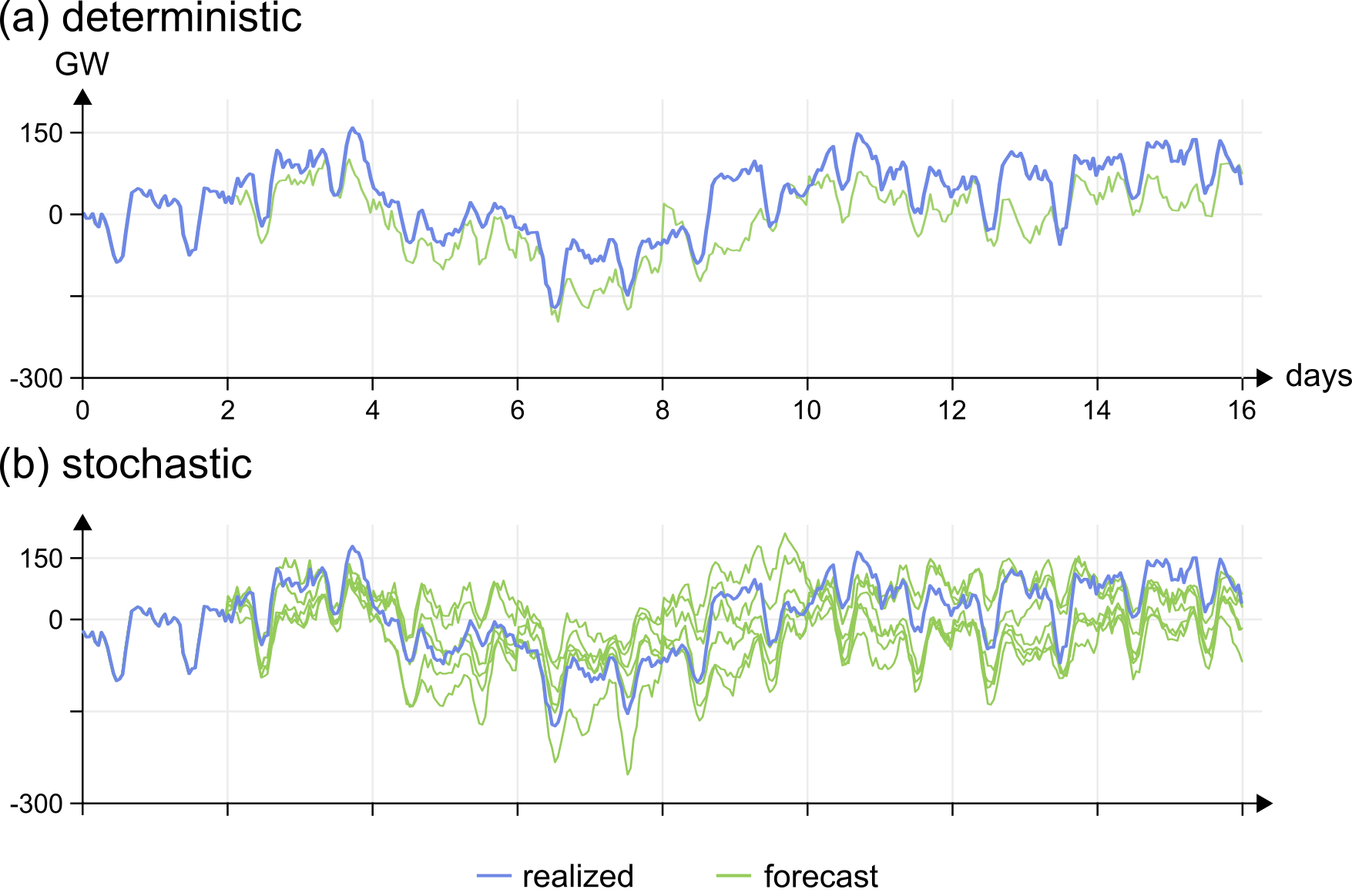}
\caption{Example forecast data for one step of the rolling horizon.}
\label{fig:curves}
\end{figure}

Fig.~\ref{fig:curves} shows an example of the input data for one step of the rolling horizon, with the realized time series as a reference. While deterministic dispatch uses a single forecast, stochastic dispatch uses a range of forecasts fitted to a scenario tree. The range of forecasts fans out until the sixth day but remains constant afterward, because the fitting algorithm prioritizes early branching.

\section{Benefit of multi-day stochastic dispatch} \label{sec:res2}

This section presents the results of the rolling-horizon benchmark, first for a reference case with a 90\% share of variable renewables and then for several sensitivities. The analysis excludes renewable shares above 90\%, since previous research indicates that such shares require cross-sectoral storage with synthetic fuels over time frames longer than two weeks, which is beyond the scope of this study \cite{Schill2020, Goeke2025}.

Although the optimization minimizes total operating costs, the analysis focuses on thermal generation and the share of expected energy not served (EENS), since these results are more robust. Total costs are highly sensitive to the assumed value of lost load, and estimates in the literature vary widely, from about 1,500 to 22,940~\euro/MWh \cite{CEPA2018}. Focusing on thermal generation and EENS also omits the value of the energy stored at the end of the look-ahead horizon, but this omission is negligible, since this component consistently accounts for less than 0.05\% of total costs.

\subsection{Reference case with 90\% share of variable renewables}

\begin{figure}[!t]
\centering
$\vcenter{\hbox{\includegraphics[scale=1.0]{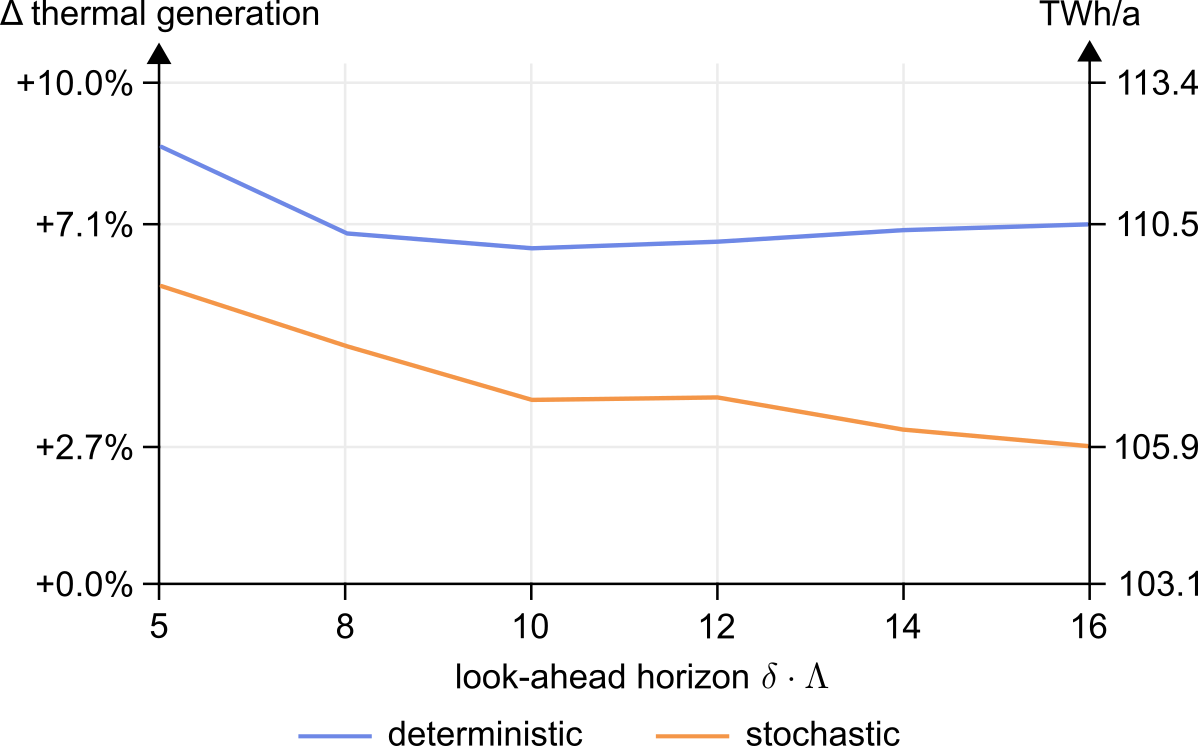}}}$
\hspace{0.0em}
\raisebox{4pt}{$\vcenter{\hbox{\footnotesize
\setlength{\tabcolsep}{3pt}
\begin{tabular}{cc}
\hline
$\delta \cdot \Lambda$ & \begin{tabular}{@{}c@{}}det. \\ EENS\end{tabular} \\
\hline
6  & 0.038 \textperthousand \\
8  & 0.045 \textperthousand \\
10 & 0.062 \textperthousand \\
12 & 0.064 \textperthousand \\
14 & 0.071 \textperthousand \\
\textbf{16} & \textbf{0.080} \textperthousand \\
\hline
\end{tabular}}}$}
\caption{Relative change in thermal generation compared with perfect foresight (left), and EENS of the deterministic dispatch (right) for the 90\%-renewables scenario. EENS of stochastic dispatch is consistently zero.}
\label{fig:impDay}
\end{figure}

For the reference case, Fig.~\ref{fig:impDay} compares the deterministic and stochastic setups, varying the look-ahead horizons, which correspond to the product of the days per stage $\delta$ and the number of stages $\Lambda$. In the left plot, thermal generation under perfect foresight serves as the optimal baseline against which deterministic and stochastic rolling-horizon dispatch are compared.

Overall, stochastic dispatch consistently outperforms deterministic dispatch, achieving lower thermal generation and zero EENS in all cases. Stochastic dispatch benefits from longer look-ahead horizons, but with diminishing returns, as forecast accuracy declines with lead time. Deterministic dispatch does not benefit from look-ahead horizons beyond 8 days. After this threshold, shorter horizons achieve similar thermal generation with lower EENS; thus, the value assigned to the final storage levels appears to guide storage operation more effectively than the increasingly inaccurate forecasts.
 
\begin{figure}[!t]
\centering
\includegraphics[scale=1.0]{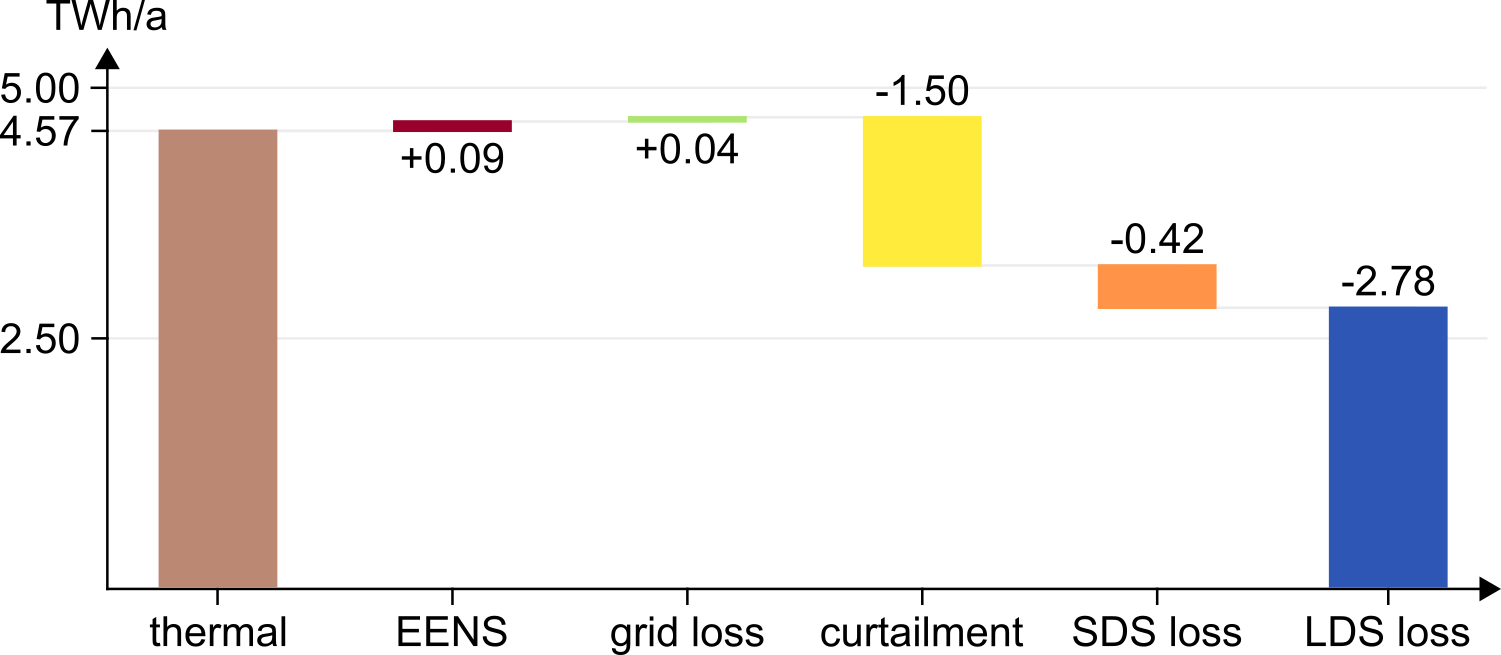}
\caption{Composition of the reduction in thermal generation achieved by stochastic dispatch for the 90\%-renewables scenario.}
\label{fig:waterfall}
\end{figure}

The waterfall diagram in Fig.~\ref{fig:waterfall} explains how stochastic dispatch reduces expected thermal generation compared to the deterministic case, as an example for a look-ahead horizon of 16 days. While changes in EENS and grid losses are negligible, most of the reduction comes from avoided curtailment and lower storage losses, especially in LDS. Curtailment decreases because stochastic dispatch better anticipates surpluses and frees up storage capacity in advance by shifting energy to other zones. Storage losses decrease because energy is stored directly in the zone where a deficit may occur, rather than first being stored in another zone and later transferred.

\subsection{Impact of forecast accuracy, renewable share, and long-duration storage} \label{sec:res2b}

\begin{figure}[!t]
\centering
$\vcenter{\hbox{\includegraphics[scale=1.0]{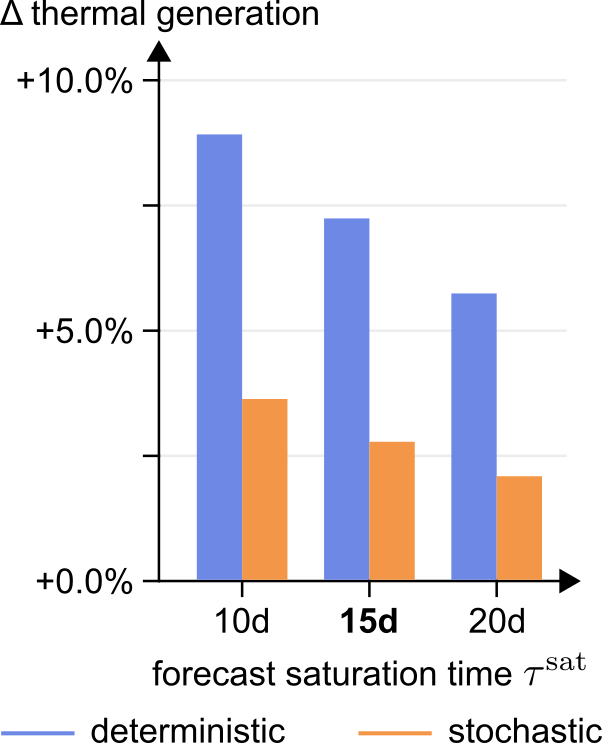}}}$%
\hspace{1.0em}%
\raisebox{4pt}{$\vcenter{\hbox{\footnotesize
\setlength{\tabcolsep}{3pt}
\begin{tabular}{cc}
\hline
$\tau^{\text{clim}}$ & \begin{tabular}{@{}c@{}}det. \\ EENS\end{tabular} \\
\hline
10 & 0.14 \textperthousand \\
\textbf{15} & \textbf{0.08} \textperthousand \\
20 & 0.07 \textperthousand \\
\hline
\end{tabular}}}$}
\caption{Relative change in thermal generation compared with perfect foresight (left), and EENS of the deterministic dispatch (right) for the 90\%-renewables scenario and different forecast accuracies. EENS of stochastic dispatch is consistently zero.}
\label{fig:accuracy}
\end{figure}

While the previous results assume average predictability of residual load, with a saturation time of 15 days until the forecast error reaches 95\% of its climatological level, Fig.~\ref{fig:accuracy} tests saturation times of 10 and 20 days. A shorter saturation time means faster convergence to the climatological error, and thus lower forecast accuracy.

For both rolling-horizon setups, the gap to the perfect-foresight solution decreases approximately linearly with the saturation time. The impact is substantial: with a saturation time of 10 days, thermal generation exceeds the perfect-foresight solution by 8.8\% under deterministic dispatch and by 3.6\% under stochastic dispatch; with 20 days, the gap shrinks to 5.7\% and 2.1\%, respectively. EENS still occurs only under deterministic dispatch and barely decreases when the saturation time increases from 15 to 20 days.

\begin{figure}[!t]
\centering
$\vcenter{\hbox{\includegraphics[scale=1.0]{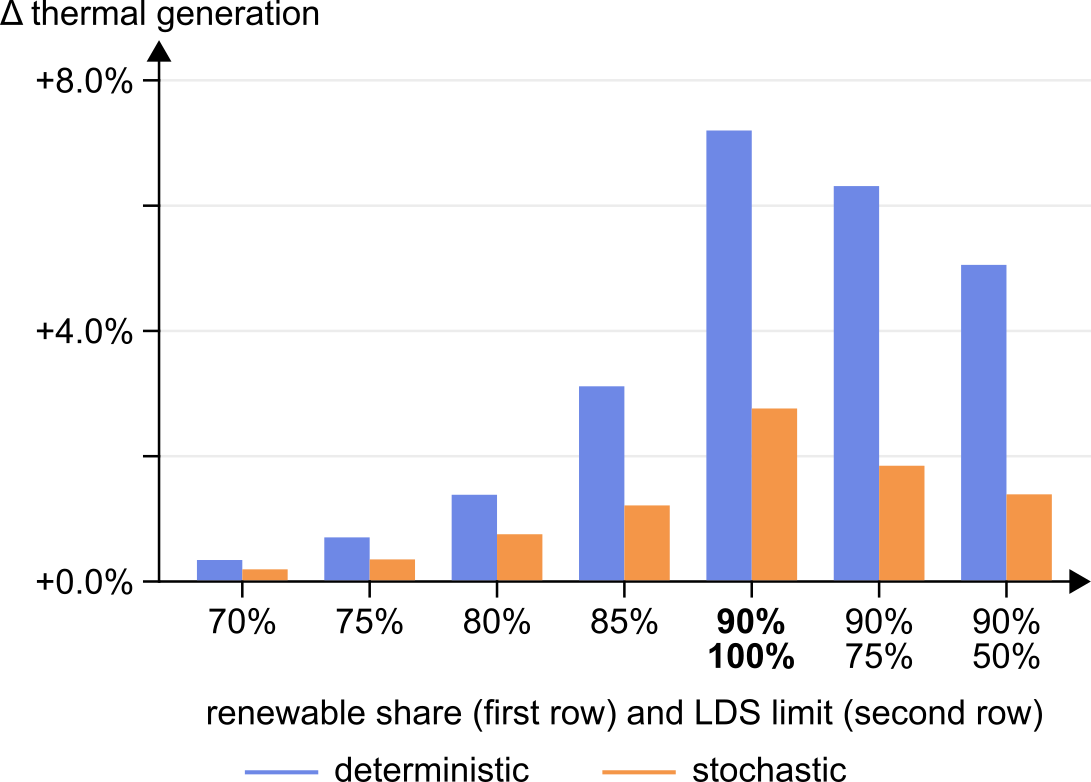}}}$%
\hspace{1.0em}%
\raisebox{4pt}{$\vcenter{\hbox{\footnotesize
\setlength{\tabcolsep}{3pt}
\begin{tabular}{ccc}
\hline
RE & LDS & \begin{tabular}{@{}c@{}}det. \\ EENS\end{tabular} \\
\hline
70 &  & 0.001 \textperthousand \\
75 &  & 0.003 \textperthousand \\
80 &  & 0.017 \textperthousand \\
85 &  & 0.013 \textperthousand \\
\textbf{90} & \textbf{100}  & \textbf{0.080} \textperthousand \\
90 &  75 & 0.019 \textperthousand \\
90 &  50 & 0.006 \textperthousand \\
\hline
\end{tabular}}}$}
\caption{Relative change in thermal generation compared with perfect foresight (left), and EENS of the deterministic dispatch (right) for different renewable shares and limits on the energy capacity of LDS. EENS of stochastic dispatch is consistently zero.}
\label{fig:eeShare}
\end{figure}

Fig.~\ref{fig:eeShare} shows the same benchmark for different configurations of the underlying energy system, again with a saturation time of 15 days. The first row of labels in the plot and the first column of the table show the share of variable renewables enforced in the capacity planning step. The second row and column give a limit on LDS capacity, expressed relative to its unconstrained optimum. This limit is also imposed in the capacity planning step and is tested only for the 90\% renewable case.

The results show that the gap in thermal generation grows exponentially with the renewable share but only linearly with the LDS capacity. The 85\% renewable case and the 90\% renewable case with a 75\% LDS limit both have an LDS capacity of 3.3~TWh, yet their results differ substantially. In the 85\% renewable case, thermal generation under deterministic dispatch is only 3.1\% above the perfect-foresight solution, compared with 7.1\% in the reference case. With the 75\% LDS limit and an unchanged renewable share of 90\%, the gap only narrows to 6.3\%.

Stochastic dispatch still prevents EENS, whereas deterministic dispatch already reaches an EENS of 0.017\textperthousand{} at a renewable share of 80\%. Limiting LDS, and thus the dependence on storage, greatly reduces EENS under deterministic dispatch, from 0.080\textperthousand{} without a limit to 0.006\textperthousand{} with a 50\% limit, since capacity expansion then relies more on thermal backup and short-duration storage. A common industry standard is an EENS between 0.01\textperthousand~and 0.02\textperthousand~\cite{usde}. In practice, however, these targets must also cover causes of lost load other than storage mismanagement, such as grid outages.

\section{Conclusions}  \label{sec:cl} 

This paper introduces multi-day stochastic dispatch for power systems that rely on LDS to balance high shares of variable renewables. To benchmark multi-day stochastic dispatch against deterministic alternatives, we propose a diffusion-based model that emulates residual load forecasts.

Our benchmark shows that stochastic dispatch consistently outperforms deterministic dispatch, reducing thermal generation and eliminating loss of load. In contrast to deterministic dispatch, stochastic dispatch benefits from a long look-ahead horizon rather than assigning a fixed value to stored energy, although with diminishing returns as forecast accuracy declines with lead time. Higher forecast accuracy generally reduces the benefit of stochastic dispatch and the overall difference to the perfect-foresight solution.

The benefits of stochastic dispatch increase with both the share of variable renewables and LDS capacity, but the effect of the renewable share is stronger. The reduction in thermal generation compared to perfect foresight is marginal at a 70\% share of wind and solar but substantial at 90\%, even when LDS capacity is restricted. Moreover, stochastic dispatch eliminates the loss of load that occurs under deterministic dispatch, which is already close to common reliability standards at a renewable share of 80\%.

The observed benefits warrant further research on multi-day stochastic dispatch in three areas:
\begin{enumerate}
\item \textbf{Forecasting:} Follow-up work with a stronger meteorological focus should develop true probabilistic multi-day forecasts as a basis for implementing stochastic dispatch in practice. Such work would also show which forecast accuracy is realistically achievable.
\item \textbf{Optimization:} Future work should investigate modeling techniques and solution algorithms that scale to larger systems and further narrow the gap to the perfect-foresight solution, since this work is limited to solving the deterministic equivalent for three zones and still finds a substantial gap to the perfect-foresight solution. 
\item \textbf{Integration:} For application in renewable energy systems, multi-day stochastic dispatch must extend beyond the power sector to sector-coupled systems, covering LDS of heat and synthetic fuels. Integration into operational practice further requires interfaces to other models, most importantly day-ahead dispatch models.
\end{enumerate}

\section*{Acknowledgments}
I would like to thank Nina Effenberger and my colleagues at the Reliability and Risk Engineering Laboratory for their input on this work.

\printbibliography
%

\newpage

 


\vfill

\end{document}